# *Colloidal Gels Tuned by Oscillatory Shear*

Esmaeel Moghimi[1], Alan R Jacob[1], Nick Koumakis[2] and George Petekidis*[1]

[1] FORTH/IESL and Department of Material Science and Technology, University of Crete, GR-71110, Heraklion, Greece
[2] School of Physics and Astronomy, University of Edinburgh, Mayfield Road, Edinburgh EH9 3JZ, UK

# Abstract

We examine microstructural and mechanical changes which occur during oscillatory shear flow and reformation after flow cessation of an intermediate volume fraction colloidal gel using rheometry and Brownian Dynamics (BD) simulations. A model depletion colloid-polymer mixture is used, comprising of a hard sphere colloidal suspension with the addition of non-adsorbing linear polymer chains. Results reveal three distinct regimes depending on the strain amplitude of oscillatory shear. Large shear strain amplitudes fully break the structure which results into a more homogenous and stronger gel after flow cessation. Intermediate strain amplitudes densify the clusters and lead to highly heterogeneous and weak gels. Shearing the gel to even lower strain amplitudes creates a less heterogonous stronger solid. These three regimes of shearing are connected to the microscopic shear-induced structural heterogeneity. A comparison with steady shear flow reveals that the latter does not produce structural heterogeneities as large as oscillatory shear. Therefore oscillatory shear is a much more efficient way of tuning the mechanical properties of colloidal gels. Moreover, colloidal gels presheared at large strain amplitudes exhibit a distinct nonlinear response characterized largely by a single yielding process while in those presheared at lower rates a two step yield process is promoted due to the creation of highly heterogeneous structures.

# I. Introduction

Industries related to food, ceramics, personal care products and tissue engineering require specific end products which are obtained through precision processing techniques, employing significant resources. Hence over the last few years, the ability to engineer structural and mechanical properties of soft matter through processing has attracted a great amount of interest. A model soft matter system can give valuable insights into the underlying mechanisms related with structural changes during processing.

The simplest soft matter model system is suspensions of colloidal hard spheres[1] which can be used as model systems, to address many fundamental phenomena of condensed matter physics such as the equilibrium phase transitions and frustrated out-of-equilibrium states such as glasses and gels transitions. [2]

Colloids with repulsive or attractive interactions have a rich and well studied thermodynamic phase diagram involving "gas", liquid and crystal phases. In addition to thermodynamic phases, kinetically arrested states start intervening with thermodynamic equilibrium at highly concentrated or strongly attractive systems.[3, 4]

Therefore while at high particle volume fractions particles are trapped entropically in cages formed by their neighbors creating an amorphous glassy system,[3] in the presence of strong interparticle attractions free floating clusters, fractal percolating particle networks, or more concentrated cluster dominated gels[5-8] and attractive glasses[4] are formed as the particle volume fraction is increased. In both cases the system exhibits a transition from an ergodic liquid to non-ergodic solid, either by increasing the volume fraction or the attraction strength. However the network will break under high enough stresses (or strains) often in multiple steps due to the existence of spatial heterogeneities[9-15] results in flow-induced structural anisotropy[16-19] while it may also show significant thixotropy, and ageing.[20, 21]

Tuning the properties of these network gels is important in a wide range of industrial applications. Such tuning is typically carried out by altering the properties or concentrations of the particles, and by changing the characteristics of inter-particle potential information (for example range and strength of attraction) or external conditions such as temperature, pH. Another route to tune a gel is by external fields such as preshear which may provide access to microstructures that can not be achieved through thermodynamic variables.[22] Shearing these colloidal systems can produce a wide variety of structures with different mechanical properties. Depending on the shear rate enhanced cluster formation or strong bond breaking may takes place in bulk,[22-29] two-dimensional[30] and microchannel flows.[31] Such distinct shear-induced structural changes has a strong impact on the yield stress, viscoelastic moduli,[21, 22, 32-35] delayed yielding[36, 37] and collapse of the gel network.[38, 39]

Understanding the detailed microscopic mechanisms involved in such flow response can be achieved through examination of model systems with tunable interactions by a combination of experimental techniques and computer simulations.

In previous work the effect of steady shear flow on both the structural and mechanical properties was studied in a model colloid-polymer mixture where the range and strength of depletion attraction was tuned by varying the size and concentration of the polymer chain respectively. The system was examined both during steady state shear and its evolution after shear cessation, both structurally and mechanically.[22] There we found that high shear rates fully break the structure into individual particles and lead after shear cessation to strong solids with relatively homogeneous structures, whereas preshear at low rates create largely inhomogeneous structures which remain stable after shear cessation and exhibit a reduced elasticity of weaker solids.

In this paper we explore the impact of oscillatory shear flow on both structural and mechanical properties for a similar intermediate volume fraction depletion gel. The structure and mechanical properties are probed during and after flow cessation using experimental rheometry and BD simulations. We demonstrate that in comparison with steady shear flow, oscillatory shear can tune much more efficiently, causing stronger variation in the structural and mechanical properties of a colloidal gel.

The rest of the paper is organized as follows: we first present the mechanical and structural properties of the gel under shear and subsequently after flow cessation. Afterwards the effect of quenching rate, types of preshear (oscillatory or steady) and inter-particle attraction strength on the tuning of the mechanical properties of the gel by shear is studied. Finally, we examine the impact of preshear on the nonlinear, yielding behavior of colloidal gels in start-up shear flow.

## II. Materials and Methods

We used polymethylmethacrylate (PMMA) nearly hard-sphere particles stabilized by chemically grafted poly-hydro-stearic acid (PHSA) chains (≈10 nm) dispersed in octadecene, a high boiling point solvent (b.p 315 °C) to avoid evaporation. Particles have a hydrodynamic radius of R=196 nm (measured in dilute regime by dynamic light scattering) with the polydispersity around 12% which suppress crystallization. The depletion attractions implemented between the particles by adding non-adsorbing linear Polybutadiene (1, 4-addition) (PB) chains (Polymer Science Inc), with a molecular weight, $M_w$=1243300 g/mol, a polydispersity index of $M_w/M_n$=1.13 and a radius of gyration, $R_g$ =34 nm (measured by static light scattering). This implies a polymer-colloid size ratio $\xi=R_g/R$ =0.17 in dilute solution. We prepared the gel with the intermediate particle volume fraction, φ=0.44 and different polymer concentrations of $c_p/c_p^*$=0.12, 0.25 and 0.5, where $c_p^*$ is the polymer overlap concentration $c_p^* = \frac{3M_w}{4\pi N_A R_g^3}$, with $N_A$ the Avogadro number. This gives the attraction strength at contact $U_{dep}(2R)$= - 5, -10 and -20 $k_BT$, respectively according to the modified Asakura–Oosawa (AO) model.[40, 41] Note however, that in dense suspensions according to the Generalized Free Volume Theory (GFVT),[42] the effective polymer-colloid size ratio is decreased to $\xi^*$=0.14, 0.12 and 0.09, while the corresponding attraction strengths are -3.1, -4.9 and -7.4 $k_BT$, respectively. Note that in all figure captions and discussion below we report the nominal values from AO potential.

Rheological measurements were performed with an ARES-HR strain controlled instrument with a force balance transducer using homemade cone-plate geometries of diameter 25 mm, cone angle 2.7° and gap size 0.05 mm with roughness of few hundred microns which is found to be enough to avoid the wall-slip in the colloidal gels[26]. Using these serrated geometries no wall-slip effects were observed as evidenced by the absence of a stress drop at low shear rates in the flow curve in agreement with previous results[26]. Moreover no indication of slip was detected in dynamic strain sweeps or in the linear viscoelastic moduli after long time ageing due to sedimentation and gel collapse as observed before in similar gels.[11, 26] The temperature was set to T=20 °C using a standard Peltier plate and solvent evaporation was eliminated by using a solvent saturation trap, which is designed to isolate the sample from the surrounding atmosphere.

A specific shearing protocol is used to study the effect of preshear on the mechanical properties of the colloidal gels. Two slightly different approaches are followed. In the first one (termed fast quench) the gel is rejuvenated at a large strain amplitude, $\gamma_0$ =800% at a frequency $\omega$=10 rad/s until reaching steady state and then is immediately submitted to the specific strain amplitude to be studied.

In the second one (called slow quench) the gel is submitted to a reverse dynamic strain sweep with the strain amplitude lowered from large strain amplitudes ($\gamma_0$ =800%) to the desired one progressively. This allows the gel to experience all the steady state strain amplitudes before shearing to the particular strain amplitude. All experiments presented below were performed through the fast quench protocol except those where we explicitly say otherwise.

We have also performed Brownian Dynamics (BD) simulations in order to get both rheological and detailed structural information. Hard-sphere interactions in BD simulations are conducted by implementing a potential-free algorithm for hard sphere interactions[43]. Attractions were introduced with the superposition of a modified Asakura-Osawa (AO) potential[44]. The potential was calculated to be the product of the osmotic pressure and the overlap volume which changes for each pair of particles. The attraction range of $\xi = 0.1$ and attraction strength of $U_{dep}(2R) = -20\ k_BT$ was set as in previous work[22]. Near the point of contact and for a distance of $\xi_g$=0.01R, we modified the classic AO potential as to introduce a constant potential of -20 $k_BT$. The choice to implement this modification was motivated so as to approach the experimental conditions by introducing a small amount of interparticle flexibility, without modifying the basic hard sphere algorithm. The specific details of the potential have been explained in Fig. S1 in the supplementary material. Affine shear with periodic boundary conditions was applied with 30000 particles having 10% polydispersity to avoid shear-induced crystallization as experimentally observed for a similar volume fraction gel.[45]

The Brownian time in experiments, calculated using the diffusion coefficient in the dilute regime, is $t_B = \dfrac{6\pi\eta R^3}{k_B T} = 0.15\ s$ with $\eta = 4.2\ mPa.s$ the solvent viscosity.

However, the time scales in the simulations and experiments are expected to be different due to lack of Hydrodynamic Interactions (HI) in the former. In experiments, particles inside clusters are expected to be slowed down by about an order of magnitude, similarly to what occurs in a dense glass[46]. Thus we rescale the experimental Brownian time-scale $t_B$ and the non-dimensional shear rate, Pe ($=\dot{\gamma} t_B$), $Pe_\omega$ ($=\omega t_B$) and $Pe_{dep}$ by a factor of 10 throughout the figures. However, it should be noted that such simple rescaling is an approximation and is not sufficient to fully capture the effects of HI on the details of microstructure in attractive suspensions.[47]

The gel was equilibrated at rest for 100 $t_B$ before shear is applied while data under oscillatory shear were collected by averaging over one cycle of oscillation, after a waiting period of 100 cycles, which in most measurements is enough to reach steady state. Similarly in steady shear simulations data were collected after reaching a total strain of 1000%.

## III. Results and Discussion

**Structure and stresses under shear:**

The steady state values of the storage and loss modulus, G' and G'', are shown in Fig. 1 as a function of the strain amplitude, $\gamma_0$, for both experiments and BD simulations. The slices of the microstructure from BD simulations are also depicted to allow direct comparison between stresses and microstructural changes under oscillatory shear. The color of the particles in these slices indicates the number of bonded neighbouring particles which varies from least bonded, blue, to highly bonded, red. In previous work it has been found that BD is able to successfully capture qualitatively the fundamental microstructural changes under steady shear flow and also after flow cessation, even though hydrodynamic interactions (HI) are not included.[22] In both experiments and BD simulations (Fig. 1) the first maxima of G'' appears at strain amplitude of about 2.5%. This maximum reflects the first yielding process and has been related to the disconnection of large weakly bonded clusters.[10, 11, 13, 14] As the strain amplitude increases further, a crossover of G' and G'' is observed in both experiments and BD simulations beyond which the sample exhibits a liquid like response (G''>G'). The slice of the microstructure in this regime ($\gamma_0$=10%) which is beyond the yielding point, indicates formation of denser clusters with larger voids under shear compared to that of the quiescent state. This cluster densification under shear leads to a higher average number of bonds as is observed from the image of microstructure (Fig 1). The second peak of G'' reflecting the second yielding process due to the intra-cluster bond breaking[10, 11] occurs in experiments around $\gamma_0$=50%. However, this second peak is not detected in BD simulations, although a small change of slope in G' is seen in this regime. This might be due to the lack of hydrodynamic interactions in BD simulations that may affect details in microstructural changes under shear as will be discussed in more detail below. Note that such discrepancy with experiments is persistent irrespective of the pre-shear protocol followed in BD simulations. The snapshot of the microstructure at the strain amplitude of $\gamma_0$=50% where the second peak of G'' is observed in the experiments indicates, as a gross finding, the creation of even more heterogeneous structure with larger clusters/voids under shear compared to those detected at the smaller strain amplitude of $\gamma_0$=10%. However, as BD images show at much larger strain amplitudes ($\gamma_0$>100%) cluster disintegration under shear takes place. Such rich strain dependent structural changes can be quantified by a modified Peclet number, $Pe_{dep}$, which considers the competition between shear and attractive forces as introduced by Koumakis et al.,[10, 22] and Martys et al.[48] and defined latter as M' by Eberle et al.[49] and Kim et al.[50] For oscillatory shear this makes sense above the yield strain amplitude where we have:

$$Pe_{dep} = \frac{F_{shear}}{F_{dep}} = \frac{12\pi\eta\xi R^3}{U_{dep}(2R)/2\xi} \frac{\gamma_0 \omega}{2\xi} = \frac{12\pi\eta\xi R^3}{U_{dep}(2R)} \gamma_0 \omega$$

where the ratio of $Pe/Pe_{dep} = U_{dep}(2R)/2k_B T\xi$ depends only on the details of the attractive potential. For the values of $Pe_{dep} > 1$, bonds between particles are expected to be disrupted by shear forces and attractive forces are essentially inactive as the gel exhibits liquid like behavior. For $Pe_{dep} < 1$, however, the system is strongly affected by inter-particle attractions with shear-induced rearrangements at the presence of attraction leading to the formation of compact clusters and hence increase the heterogeneity and voids (or clusters) compared to the quiescent state.[22]

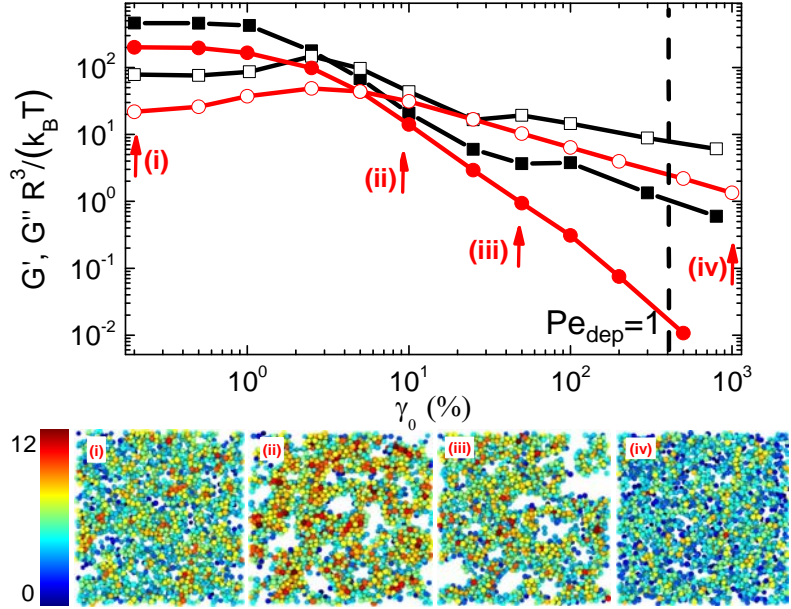

Fig. 1: Storage modulus G' (solid symbols) and loss modulus G'' (open symbols) as a function of strain amplitude $\gamma_0$ at $Pe_\omega=15$ for experiments (black squares) and $Pe_\omega=10$ for BD simulations (red circles). The corresponding images of structure with a thickness of 4R under shear taken from BD simulations are shown at the bottom. Particles are colored by the number of bonds. The value of $Pe_{dep}=1$ is shown by the vertical black line. For experiments, $\varphi=0.44$, $U_{dep}(2R) = -20k_BT$, $\xi = 0.17$ ($Pe/Pe_{dep} = 59$) and simulations, $\varphi=0.44$, $U_{dep}(2R) = -20k_BT$, $\xi = 0.1$ ($Pe/Pe_{dep} = 100$).

    The structural heterogeneity under oscillatory shear is monitored in BD simulations using a real space variable measuring the size of the inhomogeneities, called Void Volume (VV) and defined as the volume of empty space (voids) in the gel similar to previous studies[51] (see Fig. S2 in supplementary material). Fig. 2a shows the average void volume, <VV>, versus the strain amplitude of preshear at different frequencies, thus different $Pe_\omega$. It shows a non-monotonic response with strain amplitude for all $Pe_\omega$. At low strain amplitudes the average void volume first increases with $\gamma_0$ indicating cluster densification under shear and reaches its maximum value at strain amplitudes, between $\gamma_0=25\%$ and $\gamma_0=100\%$ depending on $Pe_\omega$ (first shear regime). As the strain amplitude is increased further, <VV> starts to decrease but remains still larger than the quiescent state <VV> suggesting that cluster densification under shear still takes place (second shear regime). Finally, at higher $\gamma_0$, <VV> crosses below the quiescent value signifying the onset of shear induced cluster disintegration (third shear regime), as seen by 2D images of microstructure in Fig. 1. Therefore, three regimes of shearing are identified under oscillatory shear where in the first and second regimes clusters densify under shear while in the third regime they start to break up to smaller pieces. Moreover, the creation of heterogeneous structures with enhanced size of clusters/voids is favoured at lower frequencies, where the size of clusters increases by a factor of about 150 (at $Pe_\omega=1$) as opposed to the highest frequency ($Pe_\omega=100$) where it increases only by 5 times (Fig. 2a).

    The two first regimes can be distinguished from the third one by looking at the <VV> versus Pe (Fig. 2b). Here we also show for comparison <VV> under steady shear flow. All the curves collapse onto each other for $Pe_{dep}>1$ where cluster break up (third regime) starts to take place and for sufficiently high shear rates ($Pe_{dep}\gg1$) the <VV> approaches the one expected for a liquid with the same volume fraction

indicating that at such high shear rates attractive forces are inactive. However for $Pe_{dep}<1$, where cluster compactification is observed, the structure under oscillatory shear is strongly affected by $Pe_{\omega}$. This kind of restructuring occurs for $Pe_{dep}<1$ due to a competition between bond reformation and bond breaking, both affected by frequency (or $Pe_{\omega}$) of oscillatory shear. In comparison steady shear creates structural heterogeneities that decrease monotonically with Pe and in general are weaker than in oscillatory shear. Therefore the wider variety of structures can be obtained when $Pe_{dep}<1$ depending on the strain amplitude, frequency and also type of shear.

A complementary, more local measure of structural changes and heterogeneity is provided by counting the number of neighbouring particles within the attraction range, ξ defined as the bonds. The average number of bonds per particle versus strain amplitude at different frequencies is shown in Fig. 2c. It exhibits a similar non-monotonic response with the pre-shear strain amplitude as <VV>. This non-monotonic response has also been observed in lower volume fraction gels under shear[17]. However, in contrast with <VV> the maximum number of bonds under shear is detected at much smaller strain amplitudes (at $\gamma_0$=5-10% depend on $Pe_{\omega}$) where the melting of the gel takes place (where G'=G'' see Fig. 1). This discrepancy occurs since bonds give structural information on short length scales (smaller than ξ), while <VV> represents structural information averaged on all length scales. This finding suggests that onset of melting in colloidal gels takes place at strain amplitudes where the bond number exhibits a maximum representing a measure of local structural heterogeneities, at length scales of the order of the bond range. Hence the first step of yielding is relating with bond breaking or restructuring which leads to such nonmonotonic behaviour of bond number with increasing strain amplitude. Analysis of the bonds distribution under shear shows that more heterogeneous structures manifested by a broader distribution of bonds are created at $\gamma_0$=5%, where the average number of bonds gives a maximum (see Fig. S3 in supplementary material).

The average number of bonds versus Pe for both steady and oscillatory shear in the second and third regimes collapse onto each other and for sufficiently high shear rates ($Pe_{dep}>>1$) reach the value obtained in a liquid, similarly to observation for the <VV> (Fig. 2d). Note that the "bonds" determined in a liquid are only apparent and correspond to the pairs of particles with surface to surface distance smaller than 0.2R. Those apparently increase slightly at high Pe due to the build-up of a higher particle concentration in the compression direction and also some layering of particles in the flow direction, related to shear thinning, as has already been observed in BD simulations.[43]

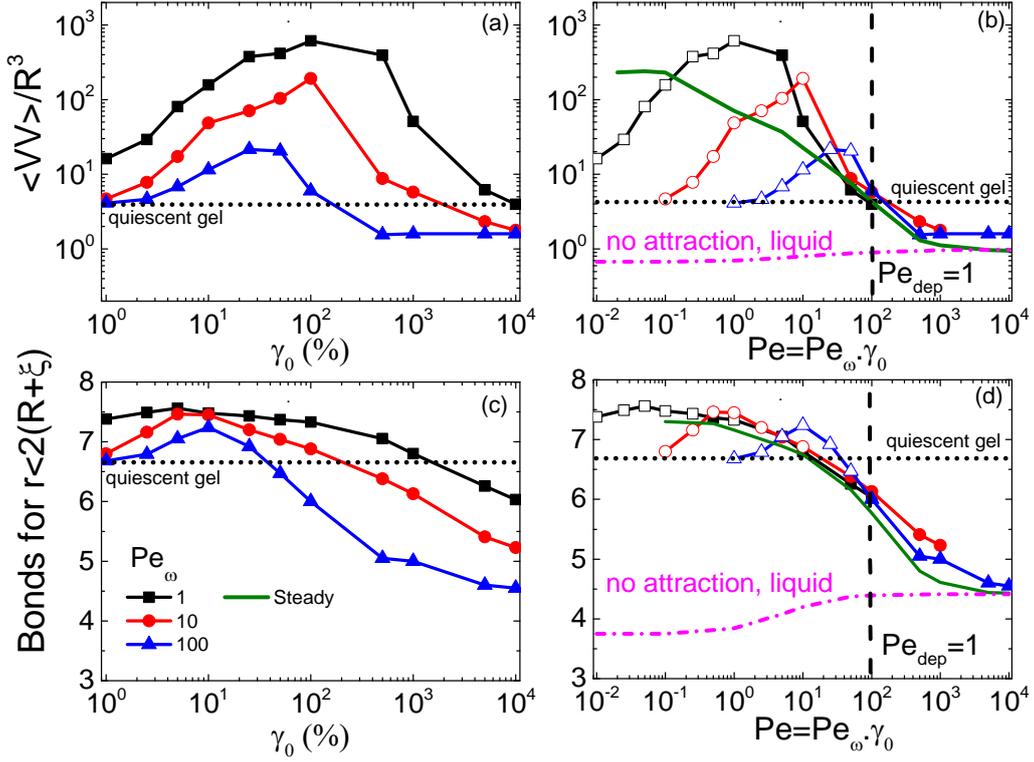

Fig. 2: Data from BD simulations at φ=0.44, $U_{dep}(2R) = -20k_BT$, ξ = 0.1 ($Pe/Pe_{dep}$ = 100). The average void volume <VV> under oscillatory shear as a function of (a) strain amplitude ($\gamma_0$) and (b) Pe at different $Pe_\omega$ as indicated. The average number of bonds per particle under oscillatory shear as a function of (c) strain amplitude ($\gamma_0$) and (d) Pe at different $Pe_\omega$ as indicated. The open symbols in (b) and (d) represent the first regime of the shear discussed in the text. The dark green curve in (b) and (d) represents results for steady shear flow. The horizontal black dotted line is the result at rest for the gel produced by quenching an equilibrated liquid. The pink curves in (b) and (d) are the result for the suspension with φ=0.44 without attraction. The value of $Pe_{dep}=1$ is shown by the vertical black dashed line.

Another way to look at the structural changes under shear is plotting the average number of bonds versus average void volume (Fig. 3). The data determined under oscillatory shear exhibit a non-monotonic dependence that is affected by the oscillation frequency (or $Pe_\omega$). By increasing the strain amplitude the average bond number initially increases with <VV> and then starts decreasing towards a maximum <VV>, beyond which both the average number of bonds and <VV> decrease together. For a constant average number of bonds, an increase of <VV> signifies the transition from a relatively homogenous structure into a heterogeneous one with larger clusters/voids and a broader bond number distribution (images 1 and 2 in Fig. 3). On the other hand, for constant <VV>, an increase of average bonds is caused by short range particle rearrangements, which lead to smaller larger scale void distribution changes (images 1 and 3 in Fig. 3).

This representation reveals again three regimes where shear affects the structure at different length scales: (i) For strain amplitudes up to about 20% the weak oscillatory shear causes short range rearrangements that increase the number of bonds but do not change the structure significantly. In this regime the increase of average bond number may be viewed also as shear-induced over-aging, i.e. shear helps the system to evolve faster than what it should under quiescent conditions. (ii) For strain amplitudes around 20%-100%, oscillations of intermediate amplitude break more bonds from initial particle network causing it to collapse and form larger clusters/voids. (iii) For even larger strain amplitudes (>100%) the strain becomes large enough to hinder large cluster formation and starts breaking clusters into smaller pieces. Moreover there is a clear frequency dependence with higher frequencies limiting cluster sizes, or the maximum <VV> to smaller values. In contrast steady shear creates clusters which are size-limited by the shear rate i.e. high shear rates results in single particles with no obvious non-monotonic behaviour.

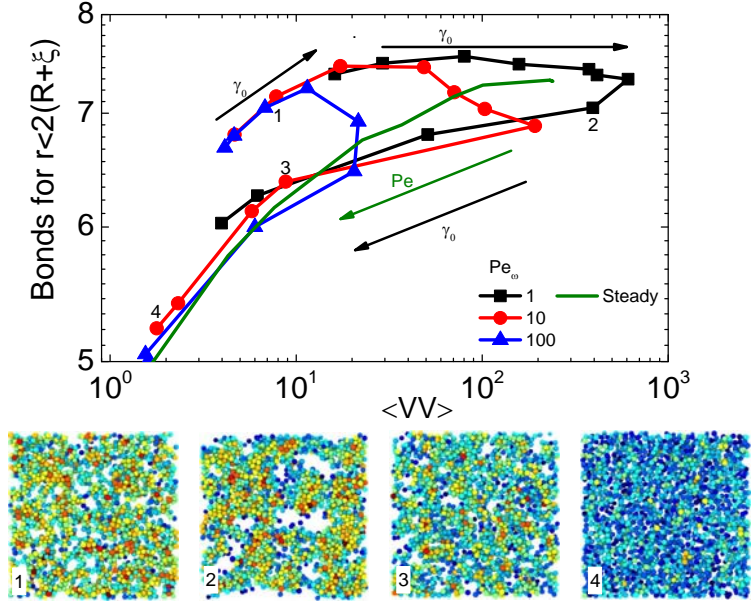

Fig. 3. Data from BD simulations at $\varphi=0.44$, $U_{dep}(2R) = -20k_BT$, $\xi = 0.1$. Average number of bonds per particle versus average void volume <VV> under oscillatory shear of different frequencies as indicated. The results for steady shear flow are shown by green dark curve. The arrows indicate the direction of increasing strain amplitude and Pe. The corresponding images of structure with a thickness of 4R under shear taken from BD simulations are shown at the bottom.

**Flow cessation:**

Fig. 4a shows the evolution of the elastic modulus, G', in rheometry experiments after flow cessation of a gel presheared at different strain amplitudes. There exists a strong correlation between the magnitude of preshear strain amplitude and the evolution of the mechanical properties of the gel after flow cessation. Fig. 4b shows G' at 5 and 600 $t_B$ after flow cessation as a function of preshear strain amplitude. After flow cessation of large strain amplitudes, G' increases strongly with time and leads into the formation of a stronger solid. Preshearing at intermediate strain amplitudes ($25<\gamma_0<800\%$), creates weaker gels which also evolve weakly with time. However, shearing the gel at lower strain amplitudes ($\gamma_0<25\%$) leads to formation of an even stronger solid which evolves even weaker with time. The weakest gel is created by preshear at $\gamma_0=25\%$.
BD simulations can be contrasted with experimental findings providing in additional valuable information on the microstructural evolution of the gel, although HI are not included. In Fig. 4 b we plot the elastic modulus deduced from BD simulations by a low amplitude oscillatory shear in the linear regime, similarly with experiments, at 1 and 150 $t_B$ after flow cessation. Here, in contrast to experiments, the G' seems to be essentially unaffected by the preshear amplitude at all times after shear is stopped, although the microstructure is quite different both during preshear and after its cessation. Microscopic images taken from BD simulations at 100 $t_B$ after shear cessation are shown in Fig.4c. After shear rejuvenation under large strain amplitudes the gel shows a strong structural evolution from a relatively homogenous liquid-like system to an interconnected gel network. However for low preshear strain amplitudes, the gel keeps the structure obtained under shear with the largest heterogeneity related to the preshearing at intermediate strain amplitudes and less for the lower strains.

In light of the excellent agreement between shear rate dependence of the microstructure deduced from confocal microscopy experiments and BD simulations under steady shear[22] we expect that the microstructure produced in BD under oscillatory shear represents well the experimental one. Moreover, as will be discussed below the general structural features of the gel after different oscillatory pre-shear strains are congruent with the non-monotonic dependence of the viscoelastic moduli found in experiments. Nevertheless the latter, surprisingly, is not captured by the elastic modulus determined in BD simulations. The origin of such apparent contradiction will be discussed latter where we compare findings from experiments and computer simulations.

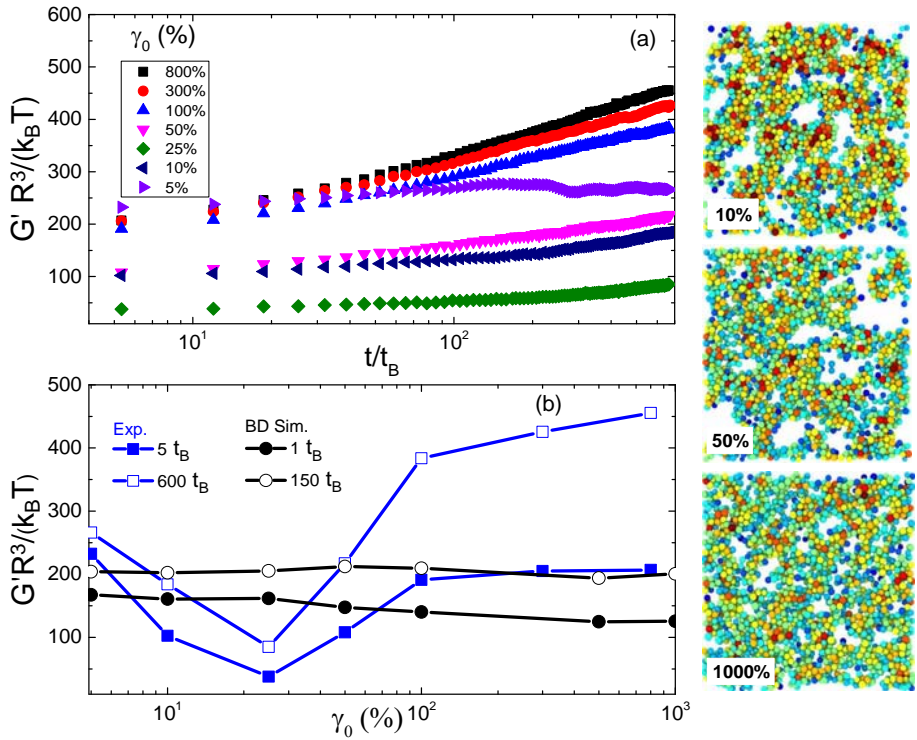

Fig. 4. (a) Time evolution of the elastic modulus, G', measured in the linear viscoelastic regime with $Pe_\omega$=15 from experimental dynamic time sweep after oscillatory shear rejuvenation at different strain amplitudes as indicated, for a sample with $\varphi$=0.44, $U_{dep}(2R)$ = -20$k_B T$ and $\xi$ = 0.17 (Pe/$Pe_{dep}$ = 59) (b) G' versus the preshear strain amplitude at different times after flow cessation as indicated from both experiments and BD simulations. For experiments data are taken from (a). (c) BD simulation images depicting the structures at 100$t_B$ after flow cessation, for a system

with φ=0.44, $U_{dep}(2R)$ = -20$k_B T$ and ξ = 0.1 (Pe/Pe$_{dep}$ = 100). Particles are colored by the number of bonds as before.

In Fig. 5 we present the full experimental linear viscoelastic response from dynamic frequency sweeps (DFS) at 600 $t_B$ after shear rejuvenation of the gel, presheared at different strain amplitudes. All samples show the typical viscoelastic behavior expected for colloidal gels with G'> G'' in the whole frequency regime indicative of solid-like response with G' increasing weakly with frequency and G'' exhibiting a minimum. The gel created by preshearing at strain amplitude of $γ_0$=25% is almost 500% weaker than the one produced by preshearing at large strain amplitudes (or equivalently through instantaneous thermal quench). As the gel elasticity was found to be a sub-linear function of polymer concentration, G~(c/c*)$^{0.9}$,[7] preparing a weaker gel with 500% lower G' would require a polymer concentration (and thus attraction strength) almost 5 times smaller. This simply indicates how efficient preshear (or in general processing) is, in tuning the linear viscoelastic properties of colloidal gels.

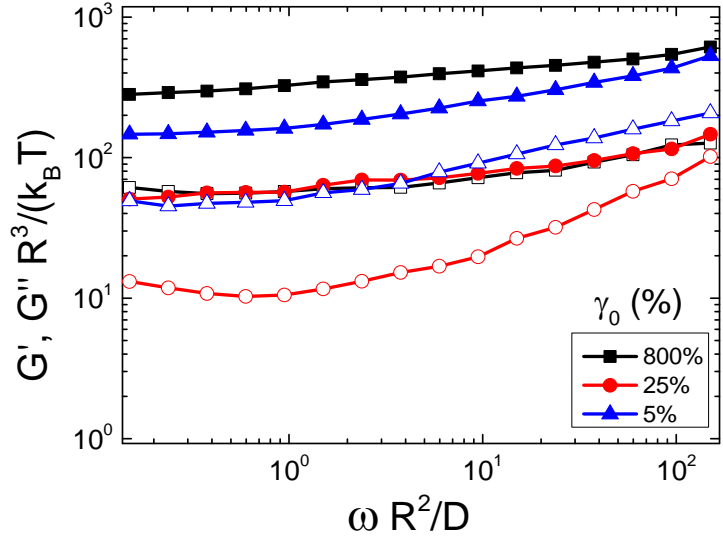

Fig. 5. Experimental dynamic frequency sweeps (with G' as solid and G'' as open symbols) performed at 600 $t_B$ after flow cessation of a gel with φ=0.44, $U_{dep}(2R)$ = -20$k_B T$, ξ = 0.17 (Pe/Pe$_{dep}$ = 59) presheared at different strain amplitudes as indicated.

The effects of preshear on the experimental elastic modulus and its time evolution (Fig. 4) can be rationalized by the impact of preshear on the structural heterogeneity. Below we discuss the viscoelastic response measured by experimental rheology in conjunction with the microstructure deduced from BD simulations. The elasticity for such intermediate volume fraction gel was found to be strongly dominated by the inter-cluster contribution.[52] There it has been proven that the elasticity is inversely related to the size of clusters and directly related to the number of bonds connecting the clusters together. Therefore gels with larger clusters and/or smaller number of inter-cluster connections will have smaller elasticity. As shown in Fig. 6a, for a gel with φ=0.44, $U_{dep}(2R)$ = -20$k_B T$ and ξ = 0.1 (Pe/Pe$_{dep}$ = 100), the average void volume, <VV>, after flow cessation has a qualitatively similar dependence on preshear strain amplitude with that detected under shear (Fig. 2). However, <VV> versus preshear strain amplitude shows an opposite trend compared to G' (Fig. 4b) which as expected confirms that the elasticity should be inversely related to the length of structural heterogeneity. For small preshear strain amplitudes, <VV> increases with $\gamma_0$ until a value of about 100% (for this $Pe_\omega$, Fig 6a), due to cluster densification. Such distinct structural changes should also have a clear signature in the mechanical properties. Therefore based of what has been explained before one would expect that the elasticity, as measured by G', after flow cessation would then exhibit a decrease with $\gamma_0$ due to both an increase of cluster size and reduction of inter-cluster connections[52]. Therefore the minimum elasticity observed in experiments can be linked with a maximum of structural heterogeneities as indicated in BD simulation by the void volume that peaks when the clusters attain their more compact form at some preshear strain amplitude. This takes place around $\gamma_0$=25% in experiments and $\gamma_0$=20-100% in BD simulations depending on $Pe_\omega$. Beyond this critical $\gamma_0$, <VV> starts decreasing with increasing strain amplitude, while the elasticity increases after flow cessation due to decreasing of the cluster size and increasing of the intercluster bonds. The time evolution of <VV> after flow cessation exhibits a qualitatively similar trend with G' measured in rheological experiments (Fig. 4). After large $\gamma_0$ preshear, <VV> evolves faster with time, as the gel quickly forms from the shear melted liquid state, and reaches the value obtained for the gel produced through instantaneous thermal quenching achieved by switching on attractions in an equilibrium liquid state, similarly with the findings for steady shear flow at high Pe.[22] In contrast, after small $\gamma_0$ preshear, there is almost no change of the <VV> with time. This is due to the fact that at low (still non-linear) preshear strain amplitudes may help the sample to approach equilibrium (liquid-gas phase separation here) introducing weak and local restructuring of the gel structure hence there is very little evolution of the structure after shear is stopped. At large strain amplitudes ($Pe_{dep}$>1), however, the system is strongly disrupted by shear and the gel is partly or fully shear melted; hence a large length scale restructuring takes place after flow cessation.

A qualitatively similar trend is observed in Fig 6b for the average number of bonds and their time evolution due to bond reformation after flow cessation. However there are significant quantitative differences, as the maximum average bond number is not detected at the same critical preshear $\gamma_0$ where <VV> exhibits a maximum. The former shows the peak at about $\gamma_0$=5% which is clearly lower than $\gamma_0$ ~100% where <VV> is maximum. As explained before this discrepancy arises from the fact that these two quantities reflect the structural information on different length scales. Bonds give information on the length scales smaller than the attraction range $\xi$ while <VV> provides information averaged on all length scales.

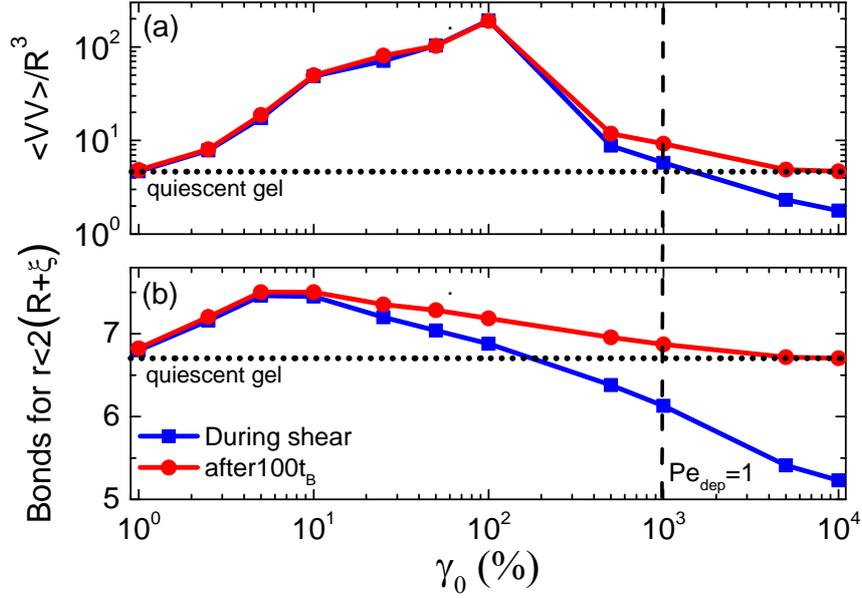

Fig. 6: BD simulations at φ=0.44, $U_{dep}(2R) = -20k_BT$, ξ = 0.1 ($Pe/Pe_{dep} = 100$): (a) The average void volume <VV> and (b) the average number of bonds per particle as a function of preshear strain amplitude ($\gamma_0$) at $Pe_\omega=10$ under shear (blue squares) and $100t_B$ after flow cessation (red circles). The horizontal black dotted line is the result at rest for the gel produced by quenching an equilibrated liquid. The value of $Pe_{dep}=1$ is shown by the vertical black dashed line.

**Effect of quenching rate:**

Fig. 7 shows the effect of the quench rate on the linear mechanical properties of the gel in experiments after flow cessation. Two types of shearing protocol are used as described in section II. Whereas in the fast quench we directly take the sample to the desired preshear strain amplitude, in the slow one, the gel is allowed to experience all steady states corresponding to higher oscillatory strain amplitudes before it gets sheared at the particular strain amplitude. For the gel presheared through the fast quench the elasticity exhibits a minimum, decreasing until preshear strain amplitude of $\gamma_0=25\%$ is reached and then increases again at higher $\gamma_0$. However for the gel presheared through the slow quench, G' after flow cessation (at 600 $t_B$, Fig. 7a) shows a monotonic increase with preshear $\gamma_0$ and for strain amplitudes larger than $\gamma_0 =25\%$ coincides with the data from fast quench. The discrepancy between two types of quenches arise for strain amplitudes less than $\gamma_0 =25\%$.

In order to elucidate the microscopic origin of such changes in the elastic properties of the gel we look into the microstructural evolution taking place during preshear of the gel with the help of BD simulations where we follow the same protocol with experiments. Fig. 7b shows <VV> as a function of preshear $\gamma_0$ for fast and slow quench in BD. The structural changes for the fast quench have already been discussed above (Fig. 2). When the gel is presheared through a slow quench, <VV> monotonically decreases with $\gamma_0$ providing a direct link with the response of the elasticity which consequently increases due to a decrease of cluster size and rise of the number of bonds between clusters. Therefore both experiments and simulations confirm that the first regime of preshearing ($\gamma_0<25\%$ in experiments and $\gamma_0<25-100\%$ in BD simulations, depending on $Pe_\omega$) is strongly affected by the exact shear history. The reason for this is that at low preshear strain amplitudes ($\gamma_0 <25\%$), but still above the first yield strain, shear is not yet exerting vast particle rearrangements and therefore the final structure (and elasticity) reached is strongly depending on the starting structure. In the case of a fast quench the starting structure is that of a gel presheared at large strain amplitudes which is relatively homogenous and so it shows large elasticity after flow cessation. However, for a slow quench the starting point is a highly heterogeneous gel which leads to the heterogeneous weak gels after shear cessation. Based on the two protocols there seems to be an optimum strain amplitude between the two yield strains (here around 100%) where the gel reaches maximum heterogeneity (and minimum G'). If the gel is sheared directly with lower strain amplitudes (as in the fast quench) the structure is not strongly perturbed and only small structural changes at local length scales are taking place leading to a strengthening of the mechanical properties. This emphasizes that not only the preshear rate is important but also the way by which shear rates are changed and therefore the way the gel is quenched has strong impact on the structure and elasticity of system after flow cessation.

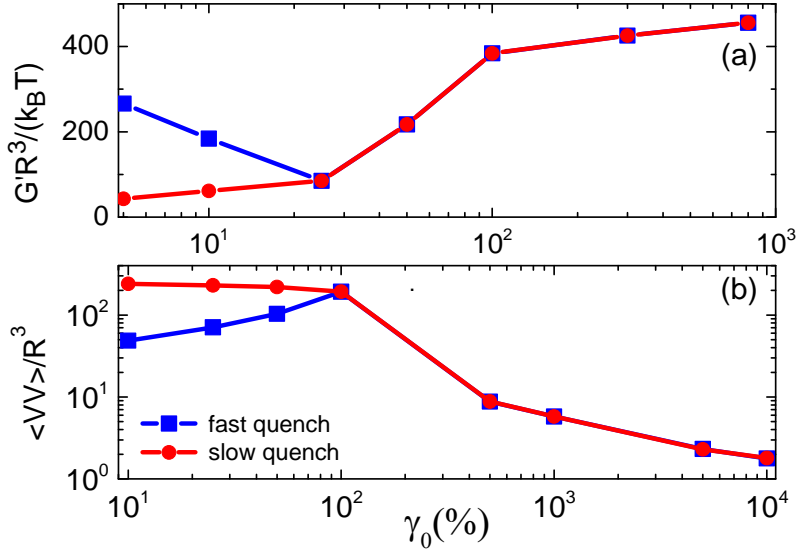

Fig. 7: (a) Experimental G' as a function of preshear strain amplitude at 600 $t_B$ after flow cessation for fast (blue squares) and slow (red circles) quench for a sample with φ=0.44, $U_{dep}(2R)$ = -20$k_B$T and ξ = 0.17 (Pe/Pe$_{dep}$ = 59) (b) The corresponding average void volume <VV> under shear taken from BD simulation for a similar system with, φ=0.44, $U_{dep}(2R)$ = -20$k_B$T and ξ = 0.1 (Pe/Pe$_{dep}$ = 100).

**Comparison of steady and oscillatory shear rejuvenation:**

Fig. 8 shows a comparison of oscillatory and steady shear rejuvenation on the elastic properties of the gel reformed after flow cessation. For both types of rejuvenation the elastic modulus after 600 $t_B$ from cessation of shear are plotted versus the dimensionless Peclet number, Pe= $\gamma_0 \omega t_B = \dot{\gamma} t_B$, allowing comparison between steady and oscillatory shear flow data. While qualitatively both types of shear lead to a weakening of the gel (lower elasticity) after an intermediate Pe preshear, quantitatively there is a large difference. A gel pre-shared in an oscillatory way at intermediate Pe's exhibits a much stronger drop in elasticity after flow cessation compared to the one rejuvenated via steady shear flow. This difference in the mechanical response is linked to the magnitude of shear-induced cluster densification at the microstructural level. Fig. 2b indeed shows that oscillatory shear is creating heterogeneous systems with structures with larger voids than steady shear flow.

Hence the stronger drop in the elasticity is a direct consequence of the creation of structures with larger clusters/voids and consequently with less number of inter-cluster connections that provide the elasticity in the network.

At the high Pe regime ($Pe_{dep}>1$), however, both steady and oscillatory preshear results into the same elasticity after flow cessation as in both cases similar structures in terms of the average void volume and the average number of bonds per particle under shear are created (see Fig. 2b and 2d). Furthermore for $Pe_{dep} \gg 1$ a complete disintegration of clusters and networks into individual particles is expected; therefore one may consider this as the regime where a proper and full rejuvenation is achieved.

These results emphasize tuning the structure in colloidal gels depends not only on the intensity of preshear (Pe) but also strongly the way that the gel is presheared. Therefore controlling the details of shear (or stirring and shaking in practical applications) is an important parameter to the final structure and mechanical strength of such systems. Within this work oscillatory shear provides a method of significantly lowering the viscoelasticity of the gel, while other shear fields might be able to widen the range of properties even more.

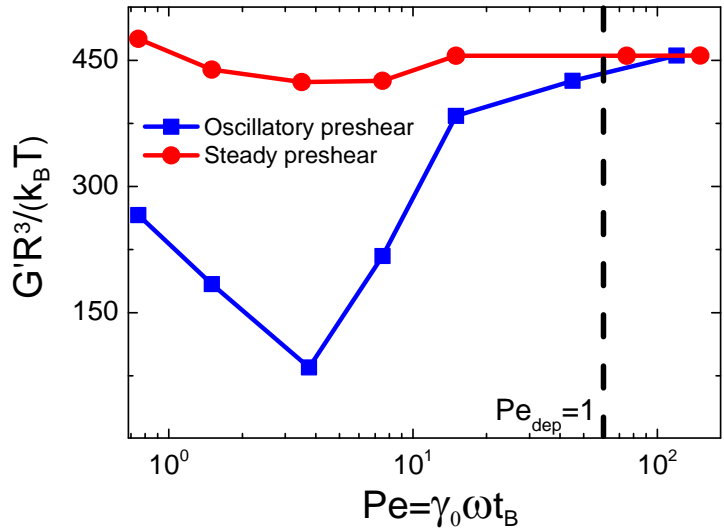

Fig. 8: Experimental storage modulus G' measured in the linear viscoelastic regime at 600 $t_B$ after flow cessation as a function of the Pe of preshear for oscillatory (blue squares) and steady (red circles) preshear. The oscillatory preshear was performed at a frequency $\omega=10$ rad/s corresponding to a $Pe_\omega=15$. The value of $Pe_{dep}=1$ is shown by the vertical black dashed line. For experiments, $\varphi=0.44$, $U_{dep}(2R) = -20k_BT$, $\xi = 0.17$ ($Pe/Pe_{dep} = 59$).

**Role of attraction strength:**

We next discuss the effect of the attraction strength. Fig. 9 shows the dependence of the storage modulus, G', measured in experiments 600 $t_B$ after flow cessation, on preshear $\gamma_0$ for three different attraction strengths, all with the same range of attraction. G' are normalized by their corresponding values after a large strain amplitude ($\gamma_0 =800\%$) preshear, allowing comparison of shear-induced mechanical changes for different systems.

The low attraction strength ($U_{dep}(2R)$ =-5 $k_BT$) gel nearly does not show any reduction of its elasticity by preshear in contrast with the higher attraction strength gels. This proves that bulk gels with stronger attractions are densified more under shear as has been shown by direct microstructural observations in two dimensional attractive systems.[53] The reason for this is that heterogeneity in quiescent gels depends on the attraction strength and exhibiting a maximum around the gelation point and decreases with increasing attraction strength inside the gel state towards some steady state value at high attractions.[7, 22] Therefore clusters in relatively more homogenous and stronger attractive gels are expected to be compactifed more when, agitated by shear, larger heterogeneities are induced as the system is driven towards the thermodynamic equilibrium i.e. phase separation. On the other hand lower attractive strength gels, already heterogeneous in nature, do not experience significant cluster compactification (or further increase in their heterogeneity) under shear and therefore their elasticity is not affected that much after flow cessation. Moreover lower attraction gels are expected to exhibit stronger aging due to Brownian activated microstructural coarsening under quiescent conditions[7, 22, 54, 55] than strong gels in which Brownian motion is entirely frozen; therefore in the latter shear-induced microstructural changes are more important (an effect of shear induced overaging) than in the former.

This finding is in agreement with arrested phase separation mechanism as stronger gels well inside the gel boundary are essentially quenched further away from thermodynamic equilibrium and thus might be assisted by shear to move towards a more stable structure closer to a phase separated system.

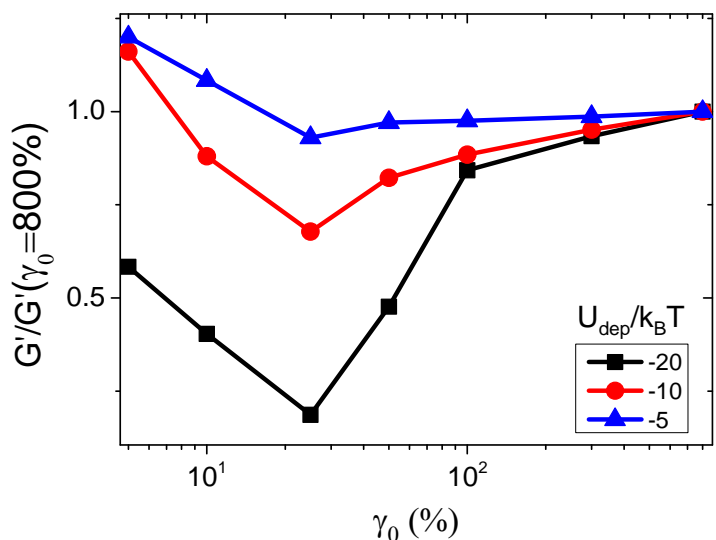

Fig. 9. Experimental storage modulus, G', measured in the linear viscoelastic regime at 600 $t_B$ after flow cessation as a function of preshear strain amplitude for different attraction strengths as indicated. For all the gels φ=0.44 and ξ = 0.17. G' is normalized by its value for the gel presheared at $\gamma_0$ =800%.

The effect of the bond strength could in principal be quantified through calculating the bond escape time as was previously done for example by Laurati et al.[11] and Smith et al.[45], where Kramer's escape approach was used to calculate the bond lifetime as a function of the strain and frequency. However, although the bond life time can be approximately estimated according to the above in order to compare it with the critical strain amplitude where the gel yields or crystallizes, its link with restructuring of dense gels structure requires more complicated analysis. Along the same lines recent work by Whitaker and Furst[56] predicts the rapture force distribution from similar Kramer's escape models showing the importance of lubrication forces for correctly predicting experimental data at the two particle level. However, extending such models to higher volume fractions require the introduction of many particle effects as well as incorporation of a bond reformation probability.

**Nonlinear dynamic response:**

Finally we examine the nonlinear, transient yield of the colloidal gels that are tuned by oscillatory shear according the procedures discussed above. As shearing these gels at different rates affects details of their microstructure, such as the length scale of clusters, voids and average bond number per particle (Fig. 2), it is reasonable to expect that not only the linear viscoelastic properties are modified, as presented above, but also their nonlinear mechanical properties are affected. Here the key point is that we may create different level of heterogeneities not by changing the interparticle potential, for example the strength of attraction,[7, 22, 54, 55] its range,[10, 57] or even the particle volume fraction,[10] but simply by changing the preshear history. Hence this allows us to directly study the impact of structural heterogeneity on the yielding response of colloidal gels, using the same samples, but simply changing the preparation protocol.

To study the nonlinear response, the gel is submitted to oscillatory shear of different shear strain amplitudes at a frequency $\omega=10$ rad/s and after shear is switched-off the gel is allowed to restructure for 600 $t_B$ (see Fig. 4). Immediately after, a step-rate experiment is applied. We performed different such start-up shear rate experiments varying the applied shear rate, and monitoring the shear stress as a function of time or accumulated strain ($\gamma$). Fig. 10 depicts start-up experiments conducted at different step rates for a gel that was presheared at different oscillatory strain amplitudes, revealing a rich response depending on the preshear conditions.

In agreement with the general picture deduced from existing studies so far in similar depletion gels in a range of volume fractions,[9-11] we see here that the gel presheared at low and intermediate oscillatory strain amplitudes ($\gamma_0<50\%$) exhibits a two step yielding process with the first yielding appearing either as a peak (after a preshear amplitude, $\gamma_0=50\%$) or reduced to a shoulder (for $\gamma_0=10\%$ and $25\%$). In contrast, the gel presheared at large oscillatory strain amplitudes ($\gamma_0>100\%$) in general exhibits a simpler response typically with a single stress overshoot indicating one yielding process. This is consistent with the physical picture of two step yielding in attractive systems related with two length scales as its existence in start-up experiments is promoted to be preshear history (here at low and intermediate oscillatory strain amplitudes) that enhances spatial heterogeneities.

Another important finding is that whereas preshearing the gel at different strain amplitudes results in a different transient response in start-up shear, the final steady state reached has an almost identical stress, indicative of unique shear-melted state where the structures that had been created during preshear at different strain amplitudes have been annealed by steady shear.

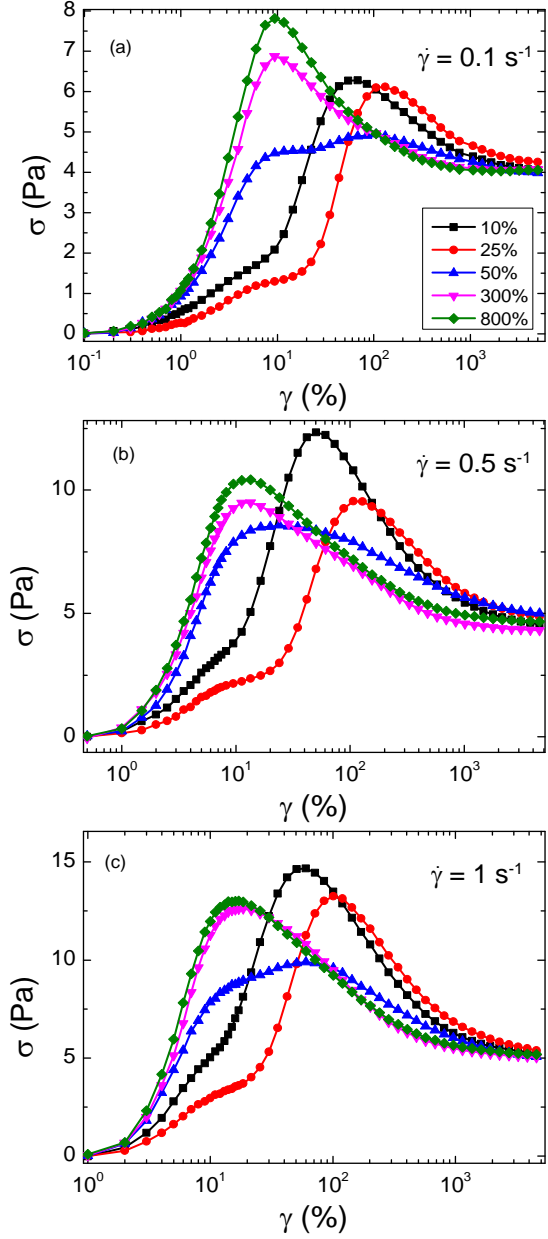

Fig. 10. Experimental step rate tests conducted on a colloidal gel at 600 $t_B$ after flow cessation of different oscillatory strain amplitudes as indicated. The test performed on a sample with φ=0.44, $U_{dep}(2R)$ = -20$k_B$T and ξ = 0.17 (Pe/$Pe_{dep}$ = 59) at steady shear rates of (a) 0.1 $s^{-1}$ (b) 0.5 $s^{-1}$ and (c) 1 $s^{-1}$.

The stress response has been associated to the shear-induced structural anisotropy through the spherical harmonic expansion of the pair distribution function and a stress-SANS rule that includes bond break-up and formation via a structure kinetics model.[58] The dependence of the yield points, as determined by the yield strain and stress values associated with the single or double stress overshoots, on preshear $γ_0$ is shown in Fig. 11. The first yield strain, $γ_1$, related typically to short length-scale bond breaking in colloidal gels,[10, 11] ranges from 5% to 15%. It increases with oscillatory preshear strain amplitude and reaches a constant value depending on the steady shear rate (Fig. 11a). In general for all different preshear histories the first yield strain is shifting to higher values with increasing shear rate of the start-up test. This increase has been attributed to the reduced opportunity for bond breaking at high rates[11], as Brownian motion becomes less effective in assisting particle escape. The first yield stress, $\sigma_1$, (Fig. 11c) measured at the first peak of the start-up shear shows a similar non-monotonic response with preshear as G' (Fig. 4b).

The second stress overshoot mainly manifested in the gels presheared at low strain amplitudes is comparable to the length scale of clusters and is related to the breaking of clusters into smaller pieces, or individual particles.[10, 11] The detections of the second stress overshoot mainly after low and intermediate preshear strain amplitudes indicates that creating large length scale heterogeneities by preshearing at low rates (Fig. 2a) promotes two step yielding.

The second yield strain, $\gamma_2$, (Fig. 11b) has a non-monotonic response with oscillatory preshear $\gamma_0$ similar to the trend observed for <VV> (Fig. 2a). The maximum value of $\gamma_2$ is found for the gel presheared at strain amplitude of 25% which suggests that the largest clusters (or voids) are created by preshearing the gel at this strain amplitude. The second yield strain decreases with increasing the imposed shear rate in start-up experiments due to decreasing the size of free clusters which are remaining after the first network rupture.[11] The corresponding stress at the second overshoot, $\sigma_2$, shows a small decrease with preshear strain amplitude (Fig. 11d) but a constant increase with steady shear rate similarly with $\sigma_1$ (Fig. 11c-d) in agreement with previous studies.[10, 11]

One can explain these findings by the effect of preshear on the structural heterogeneity (Fig. 2). As shown before, preshearing the gel at large strain amplitudes leads to the formation of relatively homogenous structures with no distinguishable clusters, at such intermediate to high volume fractions. This leads to a single yielding process during start-up which is mainly related to short length scale bond breaking. On the other hand preshearing the gel at low strain amplitudes leads to formation of highly heterogeneous structures with large distinguishable voids and rather compact clusters. This gives rise to two yielding process one related to breaking of bonds connecting the clusters and the second related to the melting of the larger clusters into smaller pieces.

Similar effects of oscillatory preshear are detected in dynamic strain sweep (LAOS) tests (see Fig. S4 in supplementary material), where performing LAOS test after oscillatory preshear with different strain amplitudes affects two-step yielding in a similar way with step rate tests. Equivalently, successive dynamic strain sweeps with increasing and decreasing strain amplitude exhibit hysteresis as they correspond to different preshear history (Fig. S5 in supplementary material).

The effects of preshear history on transient rheology and yielding of colloidal gels shown above, manifest a new route to tune the nonlinear mechanical properties of relevant products in addition to their linear viscoelasticity. Therefore systems/ products comprising of similar attractive colloids, if properly prepared can be more or less resistant to shear before they yield and flow.

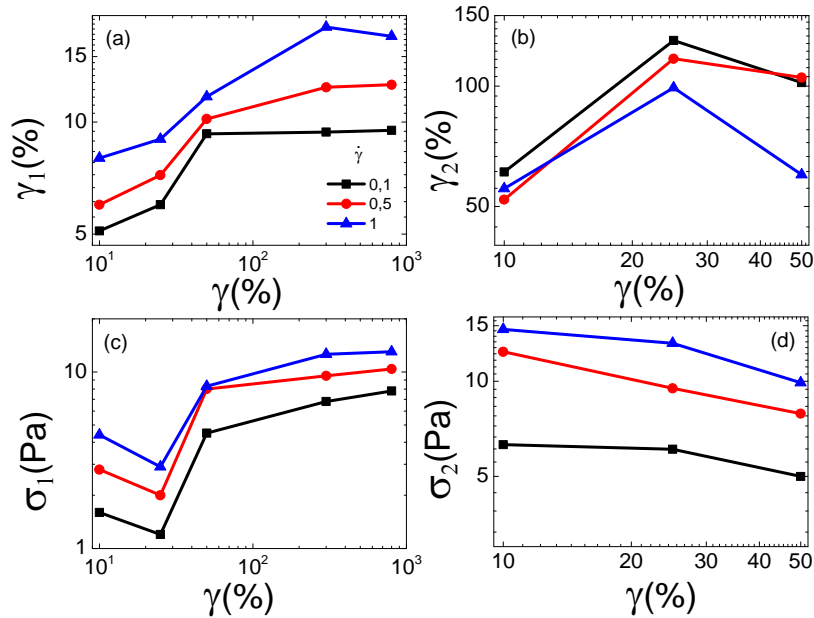

Fig. 11. (a) First yield strain (b) second yield strain (c) first yield stress and (d) second yield stress as a function of applied preshear strain amplitude, taken from Fig. 12 for different shear rates as indicated.

## IV. Discrepancies between experiments and BD simulations

As a final point, we discuss the discrepancies observed between experimental rheology and results from BD simulations and provide the possible origins for such differences. As mentioned earlier there are two main deviations. Firstly BD simulations do not capture the non-monotonic dependence of the linear elasticity of the gel (Fig. 4b) as a function of preshear strain amplitude, detected in experiments. Secondly, as has been seen in other studies[16, 59], BD simulations can not accurately produce, at such intermediate and low particle volume fractions the two step yielding that is widely seen in experiments. For the latter, since hydrodynamic interactions are absent in Brownian Dynamics simulations, they become the obvious possible origin for such deviations. At intermediate and low volume fractions where spatial heterogeneities are important, the absence of two step yielding in BD could be due to the lack of long range HI that would affect flow, deformation and rupture of clusters. To verify such mechanism though direct comparison between BD and large scale Stokesian Dynamics simulations under non-linear shear conditions are required, something which is a quite challenging task. Note that computer simulations accounting for HI already have demonstrated that the HI significantly promote gelation of attractive colloids at rest and lead to formation of more open/heterogeneous structures.[47, 60-62] On the other hand the possible origin for the lack of the non-monotonic dependence of G' after preshear in BD simulations (Fig. 4b) is still not clear.

To this end we provide some possible explanations that need to require a further studies to be verified. A first potential cause might be related with the details of the micromechanics and of the attractive potential that is approximated in BD based on the AO potential. Yet in real experimental systems, the depletion attraction is induced by real polymer chains, whose distribution inside a dense cluster of colloids is not clear. Therefore in the real experimental system, attractions within particle clusters might be different than that postulated simply based on the polymer concentration and the AO potential, or expected of other attractive systems such as sticky spheres. Along these lines, the range and depth of attractions might be reduced and eventually particles within clusters might in reality be more weakly bonded than expected. That would result in a weaker stresses in the experimental system (i.e. with lower G'), specifically showing in the intermediate strain amplitude preshear where strong cluster densification takes place.

Another possible source of this discrepancy is indeed HI. The presence of full HI in the real system is expected to affect the flow of clusters past each other. Therefore, some local structural details might be affected, although the general findings provided by current BD are in agreement with experiments as proven previously for gels under steady shear[22]. Moreover , following shear cessation, the time evolution of the structures created at different preshear amplitudes should be altered by HI in a nontrivial way depending on the cluster size, or dominant length scales, which is present in each case[55]. The above speculations should be thoroughly tested by experiments that are capable of providing detailed information on both at short and long length-scales, such as simultaneous rheometry and confocal measurements and by computer simulations where full HI are included.

Finally, we should mention that although start-up shear tests by BD simulations show only essentially one stress peak, we do see some clear effect of the response depending on the amplitude of the oscillatory preshear (see Fig. S6 in supplementary material). The effects are detected both around and beyond the stress peak with the creation of a shoulder at intermediate oscillatory preshear amplitudes where the structure is more heterogeneous. While these findings suggest that the effects of preshear are more prominent in the nonlinear response and therefore detected even with BD simulations they also provide evidence for the importance of the full hydrodynamic interactions. Moreover as shown by Park et al.,[16, 58] although the role of actual surface interactions in principle may affect the details of yielding behavior of the gel, even considering non-central attractive interactions, the double yielding is not detected in BD simulations. To this end support of the vital role of HI in the flow and yielding of colloidal gels was recently provided by Dissipative Particle Dynamics simulations at lower volume fraction gels that managed to demonstrate a clear two step yielding in start-up shear tests.[63]

## V. Conclusions

We have investigated the structural and mechanical properties of colloidal gels at intermediate volume fraction during and after shear cessation as a function of the oscillatory shear strain amplitude used during preshear. We use both experimental rheometry and Brownian Dynamics simulations to gather comprehensive information on the structural evolution and its relation with stress response and compare the oscillatory with steady preshear. Analysis of microstructure is performed by determining the average void volume, a quantitative measure of spatial heterogeneities and the average number of inter-particle bonds. Microstructural analysis reveals that a variation in the applied strain amplitude in oscillatory preshear introduces strong variations in the structure of the gel both under shear and during gel reformation after flow cessation.

Structural heterogeneities (cluster/void size) show a non-monotonic dependence with the oscillatory preshear strain amplitude. At low strain amplitudes, where the attractive forces are dominating over shear forces ($Pe_{dep}<1$), the structural heterogeneity initially increases with the strain amplitude due to shear induced cluster compactification until a strain amplitude where the maximum cluster compactification under shear takes place (first regime). Beyond this point heterogeneities decrease with increasing preshear strain amplitude but still remaining larger than those found at rest (second regime). Finally, shearing the gel at large strain amplitudes, where $Pe_{dep}>1$, produces homogenous structures due to cluster disintegration under shear (third regime).

These distinct microstructures created under shear are of central importance in the gel reformation after shear cessation, leading to materials with different final microstructures and therefore mechanical properties at longer times. Gels reform after large oscillatory strain amplitude preshearing into stronger solids with a relatively homogeneous structure. Preshearing at intermediate strain amplitudes creates gels with weaker elasticity and highly heterogeneous microstructures. Finally, under low strain amplitude preshear (still above the first yield strain though) the system is again driven into a stronger gels with relatively less heterogeneous microstructure than those created at intermediate strain amplitudes, as the weak oscillation probably causes small scale bond restructuring that promotes stronger configurations.

In comparison BD simulations indicate, that steady shear flow produces weaker structural heterogeneities compared to oscillatory preshear. This is the reason why colloidal gels may become weaker when presheared in an oscillatory manner.

Finally we demonstrate that preshear has a strong impact on the nonlinear, yielding behavior of colloidal gels. Gels that are presheared oscillatory at large strain amplitudes exhibit a largely single yielding response during a start-up test. On the other hand in those that are rejuvenated at low and intermediate oscillatory strain amplitudes (and equivalently lower shear rates) a two-step yielding process is promoted due to the increase of structural heterogeneity during preshear.

This work provides further understanding of the way colloidal gels flow under shear, especially regarding the link between the impact of preshear (or type of rejuvenation) on microstructure and the mechanical properties of the system. Therefore it might be used as a predictive tool for processing of complex soft materials with desired properties.

## Acknowledgments:

We thank A. B. Schofield for providing the PMMA particles. We acknowledge funding from Greek projects Thales ''Covisco', Aristeia II ''MicroSoft'' and EU project "SmartPro". N.K. has been supported by EU Horizon 2020 funding, through H2020-MSCA-IF-2014, ActiDoC No. 654688.